\documentclass[doublecol]{epl2}
\usepackage{amssymb}

\shorttitle{Quantum entanglement} \institute{ \mbox{Department of
Physics and Astronomy, Lehman College, City University of New
York,} \\ \mbox{250 Bedford Park Boulevard West, Bronx, New York
10468-1589, U.S.A.} } \pacs{75.80.+q}{Magnetomechanical effects,
magnetostriction} \pacs{75.45.+j}{Macroscopic quantum phenomena in
magnetic systems} \pacs{75.50.Xx}{Molecular magnets}
\pacs{85.65.+h}{Molecular electronic devices} \abstract{ We solve
Schr\"{o}dinger equation describing a tunneling macrospin coupled
to a torsional oscillator. Energy spectrum is studied for various
quantum regimes. Magnetic susceptibility and noise spectrum are
computed. We show that entanglement of the spin with mechanical
modes of a subnanometer oscillator results in the decoherence of
spin tunneling. For larger oscillators the presence of a tunneling
spin can be detected through splitting of the mechanical mode at
the resonance. Our results apply to experiments with magnetic
molecules coupled to nanoresonators. }

\begin{document}

\title{Quantum entanglement of a tunneling spin with mechanical modes of a
torsional resonator}
\author{D. A. Garanin \and E. M. Chudnovsky}
\maketitle

\section{Introduction}

There has been enormous progress in measurements of individual nanomagnets
\cite{wer07natmat}, microcantilevers and microresonators \cite
{greyurbusarn96epl,irisch03prb,lahbuucamsch04sci,walmorkab06apl,houcketal07nat,poggioetal08nat, regteuleh08nat,janmorkab01apl,mor03jpd,chamor03jap,gaoetal04jap,rugbudmamchu04nat,wangetal06prl, davisetal10apl}%
. Experiments have demonstrated that a mechanical torque induced by the
rotation of the magnetic moment may be used for developing high-sensitivity
magnetic probes and for actuation of micro-electromechanical devices. The
underlying physics is a direct consequence of the conservation of the total
angular momentum: spin plus orbital. While this side of the effect is clear,
the mechanism by which the angular momentum of individual spins gets
transferred to the rotational motion of a body as a whole has been less
understood. In a macroscopic body it involves complex evolution of
interacting spins and phonons towards thermal equilibrium. In that respect
the case of a magnetic nano- or microresonator is simpler due to the great
reduction of the number of mechanical degrees of freedom.

\begin{figure}[th]
\centerline{\includegraphics[angle=-90,width=7cm]{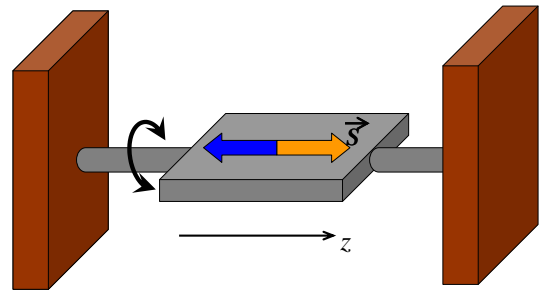}}
\caption{System studied in the paper. Macrospin (e.g., a magnetic molecule)
is attached to a torsional oscillator such that the magnetic anisotropy
(quantization) axis is parallel to the axis of mechanical rotations.}
\label{resonator}
\end{figure}

Recently, theoretical study of rotating magnetic nanosystems has been
conducted within classical \cite
{kovbaubra03apl,kovbaubra05prl,kovbaubra07prb,jaachugar09prb} and
semiclassical \cite{jaachu09prl,jaachugar10epl} approaches. When spin is
treated quantum-mechanically, further reduction of the number of degrees of
freedom can be achieved in the low energy domain where only the lowest spin
doublet that originates from the tunneling between spin-up and spin-down
states is relevant. This would be the case of, e.g., a single-molecule
magnet. Rigorous quantum-mechanical treatments have been recently suggested
for the problem of a tunneling macrospin in a freely rotating body \cite
{chugar10prb} and for the problem of a tunneling macrospin coupled to the
rotational modes of a nanoresonator \cite{kovhaybautse11prl} (see Fig. \ref
{resonator}).
These two problems have one common feature: The spin tunneling becomes
suppressed when the body containing the spin is too light. The physics
behind this effect is quite clear \cite{chu94prlconservation}.
Delocalization in the spin space that corresponds to tunneling of spin $%
\mathbf{S}$ between spin-up and spin-down states reduces the energy by $%
\Delta/2$, where $\Delta$ is the splitting of the tunneling doublet. Since
spin transitions are accompanied by the change of the angular momentum they
generate rotational motion of the body with the energy $\hbar^2{L}^2/(2I)$,
where $I$ is the moment of inertia and $L$ is the mechanical angular
momentum that is generally of order $S$. At small $I$ such rotations cost
too much energy, so that the tunneling in the ground state should be frozen.
This effect is conceptually similar to the decoherence and freezing of the
tunneling of a particle in a double-well potential due to dissipation \cite
{bramoo82prl}.

In this paper we solve Schr\"{o}dinger equation for a macrospin
coupled to a nanoresonator, Fig. \ref{resonator}. By considering
various ranges of parameters of the nanoresonator, we reproduce
previously obtained results and establish connection with the
problem of a macrospin in a freely rotating body. Qualitatively
different behavior is found for different ranges of parameters,
that can be interpreted as a quantum phase transition. We show
that the way to look for these effects is to study the
electromagnetic response of the system depicted in Fig.
\ref{resonator}. Our most important finding is that the coupling
of a tunneling spin to a mechanical resonator destroys quantum
coherence only for very small resonators -- generally resonators
consisting of just a few atoms. In resonators of greater size the
coherence is preserved, and the presence of a tunneling spin can
be detected by observing frequency splitting of mechanical
oscillations.

\section{The model}

Consider a model of a tunneling spin $S,$ projected onto the
lowest tunneling doublet, in a nanoresonator of torsional rigidity
$k$ that can rotate around the $z$-axis\cite{jaachugar10epl}, see
Fig.\ \ref{resonator},

\begin{equation}
\hat{H}=\frac{\hbar ^{2}L_{z}^{2}}{2I_{z}}+\frac{I_{z}\omega _{r}^{2}\varphi
^{2}}{2}-\frac{W}{2}\sigma _{z}-\frac{\Delta }{2}\left( e^{-2iS\varphi
}\sigma _{+}+e^{2iS\varphi }\sigma _{-}\right) .  \label{Ham}
\end{equation}
Here $L_{z}=-i\partial _{\varphi }$ is the operator of the mechanical
angular momentum, $I_z$ is the moment of inertia of the resonator, $\omega_r
= \sqrt{k/I_z}$ is the frequency of torsional vibrations, $W=2Sg\mu
_{B}B_{z} $ is the energy bias due to the longitudinal field $B_{z}$, $%
\Delta $ is the tunnel splitting of spin-up and spin-down states due to
crystal field, and $\sigma $ are Pauli matrices. As we shall see below, the
behavior of such a system depends on two dimensionless parameters:
\begin{equation}
\alpha =\frac{2\hbar ^{2}S^{2}}{I_{z}\Delta }\,, \qquad r = \frac{\omega_r}{%
\Delta}\,.  \label{alpha}
\end{equation}

In the limit of a free particle, $r=0,$ the total angular momentum of the
system with respect to the $z$ axis is conserved: $J_{z}=S_{z}+L_{z}=\mathrm{%
const}$. Tunneling of the spin changes $S_{z}$ by $2S$, and this change is
absorbed by the opposite change of $L_{z}.$ Thus tunneling occurs between
two quantum states having the same total angular momentum eigenvalue $J$.
Computation of the eigenstates of the system reduces to the diagonalization
of a 2$\times 2$ matrix. The resulting spectrum of the system has the ground
state with $J=0$ for \cite{chugar10prb}
\begin{equation}
\alpha \leq \alpha _{1}=\left[ 1-1/\left( 2S\right) ^{2}\right] ^{-1}
\label{alpha1}
\end{equation}
(heavy particle) that corresponds to the spin tunneling between up and down,
with the change in the angular momentum absorbed by the rotation of the
particle. However, for $\alpha >\alpha _{1}$ the ground state becomes
degenerate and in the limit $\alpha \gg \alpha _{1}$ (light particle) it
approaches $J=\pm S,$ which means that the spin cannot tunnel.

In the case of a particle having a restoring force, that is the subject of
this work, the total angular momentum of the spin and the mechanical
oscillator is not conserved. Conservation of the angular momentum occurs in
a larger closed system. Still, through the crystal field, tunneling of the
spin generates mechanical torque acting on the torsional oscillator \cite
{chugarsch05prb,jaachugar10epl}. This interaction can seriously reduce spin
tunneling for both small and large $r$ when the oscillator is light. In
particular, for small $r$ and large $\alpha $ (see below) the ground state
is nondegenerate but the gap between the ground state and the first excited
state becomes very small and the tunneling becomes effectively blocked. For
this problem Kovalev et al. \cite{kovhaybautse11prl} introduced another
dimensionless parameter,
\begin{equation}
\lambda =\sqrt{\frac{2\hbar S^{2}}{I_{z}\omega _{r}}}=\sqrt{\frac{\alpha }{r}
}\,,  \label{alphaKDef}
\end{equation}
that is especially useful in the case of large $r$. One of their
observations is that at $r \gg 1$ coupling of the spin to
quantized rotational vibrations renormalizes the tunnel splitting
according to
\begin{equation}
\Delta _{\mathrm{eff}}=\Delta e^{-\lambda ^{2}/2}\,.  \label{DeltaEff}
\end{equation}

To solve the quantum mechanical problem of a spin tunneling in a rotating
body, it is convenient to use the basis that is a direct product of the
two-state ``up/down'' basis for the spin and the harmonic oscillator basis
for the body. Thus we write the system's wave function $\left| \Psi
\right\rangle $ in the form
\begin{equation}
\left| \Psi \right\rangle =\sum_{m=0}^{\infty }\sum_{\sigma =\pm
1}C_{m\sigma }\left| m\right\rangle \left| \sigma \right\rangle .
\label{PsiExpansion}
\end{equation}
The coefficients $C_{m\sigma }$ satisfy the Schr\"{o}dinger equation
\begin{equation}
i\hbar \frac{dC_{m\sigma }}{dt}=\sum_{n=0}^{\infty }\sum_{\sigma ^{\prime
}=\pm 1}H_{m\sigma ,n\sigma ^{\prime }}C_{n\sigma ^{\prime }},
\label{SchrEq}
\end{equation}
where
\begin{eqnarray}
H_{m\sigma ,n\sigma ^{\prime }} &=&E_{m\sigma }\delta _{mn}\delta _{\sigma
\sigma ^{\prime }}-\frac{1}{2}\Delta _{\mathrm{eff}}  \nonumber \\
&&\times \left( \kappa _{mn}\delta _{\sigma ,-1}\delta _{\sigma ^{\prime
},1}+\kappa _{mn}^{\ast }\delta _{\sigma ,1}\delta _{\sigma ^{\prime
},-1}\right)  \label{HamME}
\end{eqnarray}
are matrix elements of the Hamiltonian, Eq. (\ref{Ham}). Here
\begin{equation}
E_{m\sigma }=\hbar \omega _{r}(m+1/2)-\left( 1/2\right) W\sigma
\label{EmsigDef}
\end{equation}
are energies in the absence of tunneling and \cite{kovhaybautse11prl}
\begin{equation}
\kappa _{mn}=\left( i\lambda \right) ^{m-n}\sqrt{\frac{n!}{m!}}%
L_{n}^{(m-n)}\left( \lambda ^{2}\right)  \label{kapmnDef}
\end{equation}
for $m\geq n$ and a similar expression with $m\leftrightharpoons n$ for $%
n\geq m,$ where $L_{n}^{(m-n)}$ are a generalized Laguerre polynomials and $%
\lambda $ is given by Eq. (\ref{alphaKDef}). In particular,
\begin{equation}
\kappa _{00}=1,\quad \kappa _{10}=\kappa _{01}=i\lambda ,\quad \kappa
_{11}=1-\lambda ^{2}.  \label{kappaLowest}
\end{equation}
For $r\gg 1$ only the ground state of the resonator is relevant in spin
tunneling because energies of all other states are too high compared to $%
\Delta $ \cite{kovhaybautse11prl}. In this case one returns to a two-state
model for the spin with the effective splitting (\ref{DeltaEff}). For small $%
r,$ the spin couples to many oscillator states and one has to diagonalize a
large matrix.

\section{Energy spectrum and static susceptibility}

\begin{figure}[t]
\par
\begin{center}
\centerline{%
\includegraphics[angle=-90,width=10cm]{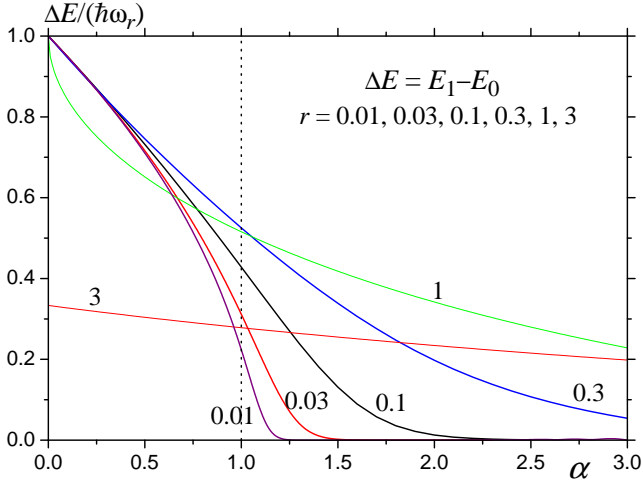}}
\end{center}
\caption{Distance $\Delta E$ between the ground state and first excited
state vs $\protect\alpha $ for different $r.$}
\label{Fig-GS_splitting_vs_our_alpha_r_fixed}
\end{figure}

Setting $C_{m\sigma }\Rightarrow C_{m\sigma }e^{-i(E/\hbar )t}$ in Eq. (\ref
{SchrEq}) results in the stationary Schr\"{o}dinger equation that can be
diagonalized numerically to find energy eigenvalues $E_{\mu }.$ The results
for the distance $\Delta E$ between the ground state and the first excited
state vs $\alpha $ for different $r$ and $W=0$ are shown in Fig. \ref
{Fig-GS_splitting_vs_our_alpha_r_fixed}. For $r\ll 1$ and $\alpha > 1$ the
ground state becomes quasidegenerate with very small, although nonzero $%
\Delta E.$ This corresponds to the localization of the spin in either
spin-up or spin-down state. On the other hand, $\Delta E$ does not
exclusively characterize the spin but also contains information about the
resonator. In particular, for $r<1$ and $\alpha \rightarrow 0$ the spin and
the resonator effectively decouple and $\Delta E\rightarrow \hbar \omega
_{r},$ which is the mode of the resonator. On the other hand, for $r>1$ and $%
\alpha \rightarrow 0$ one has $\Delta E\rightarrow \Delta ,$ which is the
spin tunneling mode.

\begin{figure}[t]
\par
\begin{center}
\centerline{%
\includegraphics[angle=-90,width=10cm]{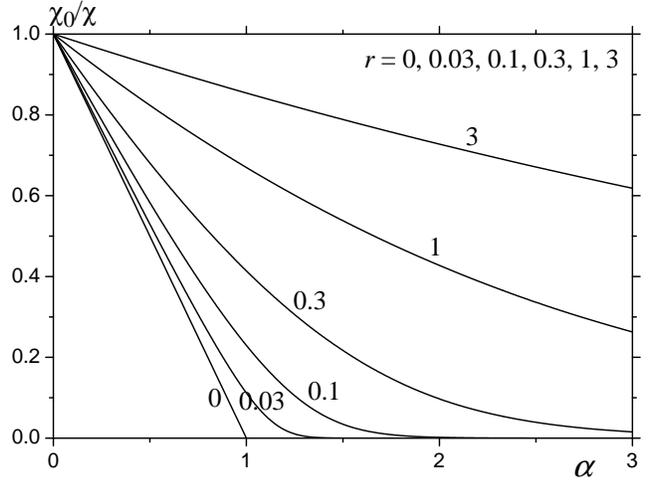}}
\end{center}
\caption{Reduced inverse susceptibility vs $\protect\alpha $ for different $%
r.$}
\label{Fig-chi_vs_our_alpha_r_fixed}
\end{figure}

The spin susceptibility is
\begin{equation}
\chi =\frac{\partial \left\langle \sigma _{z}\right\rangle }{\partial W}
\end{equation}
in the limit $W\rightarrow 0$. For a spin in a massive (non-rotating) body
one has
\begin{equation}
\left\langle \sigma _{z}\right\rangle =\frac{W}{\sqrt{\Delta ^{2}+W^{2}}}\,,
\end{equation}
thus the zero-field susceptibility is $\chi _{0}=1/\Delta .$ For a spin in a
rotating body, the effective splitting, $\Delta _{\mathrm{eff}}$, can be
defined through $\chi =1/\Delta _{\mathrm{eff}}, $ where $\chi = {\partial
\left\langle \sigma _{z}\right\rangle }/{\partial W}$ follows from the exact
numerical diagonalization of the Hamiltonian,
\begin{equation}
\left\langle \sigma _{z}\right\rangle =\sum_{m=0}^{\infty }\sum_{\sigma =\pm
1}\sigma \left| C_{0,m\sigma }\right| ^{2},  \label{chiDef}
\end{equation}
$C_{0,m\sigma }$ being the coefficients of the wave function corresponding
to the ground state, $\mu =0.$ The dimensionless ratio $\chi _{0}/\chi
=\Delta /\Delta _{\mathrm{eff}}$ vs $\alpha $ for different $r$ is shown in
Fig. \ref{Fig-chi_vs_our_alpha_r_fixed}. For $r\ll 1$ and $\alpha > 1$ the
zero-field susceptibility becomes very large because of quasidegeneracy of
the ``up'' and ``down'' spin states. For $r\gg 1$ Eq. (\ref{DeltaEff}) is
recovered.

\section{Spin-rotation resonance}

The case $\sqrt{\Delta ^{2}+W^{2}}\approx \hbar \omega _{r}$ corresponds to
the spin-rotation resonance that leads to a strong hybridization of spin and
rotational states even in the case $\lambda \ll 1.$ In the absence of the
interaction between the spin and the resonator, $\lambda =0,$ the lowest
four energy levels are
\begin{equation}
E=\left\{ \pm \frac{\sqrt{\Delta ^{2}+W^{2}}}{2},\hbar \omega _{r}\pm \frac{%
\sqrt{\Delta ^{2}+W^{2}}}{2}\right\} ,
\end{equation}
where the zero-point energy of the resonator has been dropped. The
hybridized levels are $\sqrt{\Delta ^{2}+W^{2}}/2$ and $\hbar \omega _{r}-%
\sqrt{\Delta ^{2}+W^{2}}/2.$ The truncated low-energy Hamiltonian matrix has the form
\begin{equation}
\Bbb{H}=\left(
\begin{array}{cccc}
\frac{W}{2} & 0 & \frac{\Delta }{2} & \frac{\Delta }{2}i\lambda  \\
0 & \hbar \omega _{r}+\frac{W}{2} & \frac{\Delta }{2}i\lambda  & \frac{%
\Delta }{2}(1-\lambda ^{2}) \\
\frac{\Delta }{2} & -\frac{\Delta }{2}i\lambda  & -\frac{W}{2} & 0 \\
-\frac{\Delta }{2}i\lambda  & \frac{\Delta }{2}(1-\lambda ^{2}) & 0 & \hbar
\omega _{r}-\frac{W}{2}
\end{array}
\right)
\end{equation}
where Eq. (\ref{kappaLowest}) was used. We look for $E\approx \sqrt{\Delta
^{2}+W^{2}}/2\approx $ $\hbar \omega _{r}/2$. Then for $\lambda \ll 1$ the
equation $\det (\Bbb{H}-E\Bbb{I})=0$ simplifies to
\begin{equation}
\left( E-\frac{\sqrt{\Delta ^{2}+W^{2}}}{2}\right) \left( E-\hbar \omega
_{r}+\frac{\sqrt{\Delta ^{2}+W^{2}}}{2}\right) =\frac{\lambda ^{2}\Delta ^{2}%
}{4}.
\end{equation}
At the resonance, $\hbar \omega _{r}=\sqrt{\Delta ^{2}+W^{2}},$ the
frequencies of the transition between the ground state $E_{0}=-\sqrt{\Delta
^{2}+W^{2}}/2$ and the closest excited states become
\begin{equation}
\omega _{\pm }=\frac{E-E_{0}}{\hbar }=\omega _{r}\left( 1\pm \frac{\lambda }{%
2}\frac{\Delta }{\sqrt{\Delta ^{2}+W^{2}}}\right) .  \label{splitting}
\end{equation}
This formula provides the splitting of the mechanical and spin
modes at the resonance. For such a splitting to be observable, the
quality factor of the mechanical resonator must exceed
$(1+W^{2}/\Delta ^{2})/\lambda $. Eq.\ (\ref{splitting}) can also
be obtained within semiclassical approximation \cite{jaachu09prl}.

\section{Spin dynamics}

In the problem of a tunneling spin embedded in a non-rotating crystal, the
parameter $\Delta$ has a clear physical meaning of the energy gap between
the lowest tunneling doublet. When such a spin is prepared in, e.g., the
spin-up state at $t=0$, the probability to find it in the same
state at an arbitrary moment of time $t$ oscillates on $t$
according to $\langle \sigma_z(t)\sigma_z(0)\rangle = \langle
\sigma_z\rangle_t = \cos(\Delta t/\hbar)$. When the spin is coupled
to a light oscillator and $r\gg 1$, one has to replace $\Delta \Rightarrow \Delta_{%
\mathrm{eff}}$.  At $r\ll 1$, coherent probability
oscillations are destroyed at any non-zero $\alpha$.

Spin dynamics is governed by the Schr\"{o}dinger equation, Eq. (\ref{SchrEq}%
), and the time dependence of $\left\langle \sigma _{z}\right\rangle $ is
given by
\begin{equation}
\left\langle \sigma _{z}\right\rangle _{t}=\sum_{m=0}^{\infty }\sum_{\sigma
=\pm 1}\sigma \left| C_{m\sigma }\left( t\right) \right| ^{2},  \label{sigzt}
\end{equation}
where $C_{m\sigma }\left( t\right) $ can be expanded over the eigenstates $%
C_{\mu ;m\sigma }$ as
\begin{equation}
C_{m\sigma }(t)=\sum_{\mu }a_{\mu }\exp \left( -\frac{iE_{\mu }t}{\hbar }%
\right) C_{\mu ;m\sigma },  \label{Cmsigt}
\end{equation}
the coeficients $a_{\mu }$ being determined by the initial condition. If at $%
t=0$ the spin was in the ``up'' state and the particle was in its ground
state, one has $a_{\mu }=C_{\mu ;01}^{\ast }.$ Combining these formulas
yields the time dependence

\begin{equation}
\left\langle \sigma _{z}\right\rangle _{t}=\sum_{\mu \mu ^{\prime }}A_{\mu
\mu ^{\prime }}\exp \left( i\frac{E_{\mu }-E_{\mu ^{\prime }}}{\hbar }%
t\right) ,  \label{sigtFinal}
\end{equation}
where

\begin{equation}
A_{\mu \mu ^{\prime }}=a_{\mu }^{\ast }a_{\mu ^{\prime }}\sum_{m=0}^{\infty
}\sum_{\sigma =\pm 1}C_{\mu ;m\sigma}^\ast \sigma C_{\mu^\prime;m\sigma}.
\label{Amumupr}
\end{equation}

\begin{figure}[t]
\par
\begin{center}
\centerline{\includegraphics[angle=-90,width=10cm]{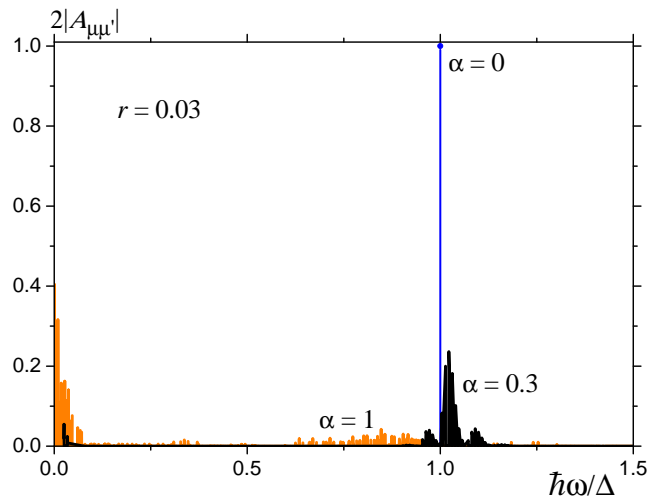}}
\end{center}
\caption{Fourier spectrum of $\langle \protect\sigma_z \rangle_t$ for $%
r=0.03 $ and $\protect\alpha =0,$ 0.3, and 1.}
\label{Fig-Resonance}
\end{figure}

Fourier spectrum of this time dependence, $2\left| A_{\mu \mu ^{\prime
}}\right| $, gives the imaginary part of the susceptibility. Plotted vs $%
\hbar \omega _{\mu \mu ^{\prime }}=E_{\mu }-E_{\mu ^{\prime }},$ it gives an
idea of the resonance absorption of the rf field by the spin. For $r\ll 1$
the Fourier spectrum consists, in general, of many lines. In the limit $%
\alpha \rightarrow 0$ the spin and the torsional oscillator decouple; in this case
only one line of height 1 remains. For $\alpha \ll 1$ there is a narrow
group of lines with a spread that gives rise to spin decoherence due to
interaction of the spin with the oscillator. At $\alpha > 1$ decoherence
becomes very strong and the low-frequency part of the Fourier spectrum
corresponding to mechanical oscillations becomes large. These results are
shown in Fig. \ref{Fig-Resonance}\ \ for $r=0.03.$

At $r\gg 1,$ there is only one line of height 1 at $\hbar \omega =\Delta _{%
\mathrm{eff}}$ with $\Delta _{\mathrm{eff}}$ given by Eq. (\ref{DeltaEff}).
This is natural because in this limit the problem is described by an
effective two-state model. For $r=1$ and small $\alpha $ there is a doublet
of lines around $\hbar \omega =\Delta $ because of the resonance interaction
between the spin and the resonator. \newline

\section{Discussion}

We have studied energy spectrum, susceptibility, and decoherence
in a system consisting of macrospin rigidly coupled to a torsional
mechanical resonator. Our general conclusion is that the coupling
does not influence spin tunneling when the resonator is
sufficiently large and heavy. However, when one approaches the
atomic size the magneto-mechanical coupling may lead to strong
decoherence of the spin states. To put these statements in
perspective, let us consider a magnetic molecule of spin $10$
embedded in a torsional resonator in the shape of a paddle of
dimensions $20\times20\times10$nm$^3$. As in Ref. \cite
{kovhaybautse11prl} we shall assume that the paddle is attached to
the walls by two carbon nanotubes of torsional rigidity
\cite{meypierot05science} $k = 10^{-18}$N$\cdot$m. The moment of
inertia of such a system is dominated by
the paddle, $I_z \sim 10^{-36}$kg$\cdot$m$^2$, so that $\omega_r = \sqrt{%
k/I_z} \sim 10^{9}$s$^{-1}$. The parameter $\lambda$ is then of order $%
10^{-2}$. For $\Delta/\hbar \ll 10^{9}$s$^{-1}$, coherent spin oscillations
at frequency $\Delta/\hbar$ will not be affected by the coupling to the
paddle. They will be more likely decohered by the coupling to nuclear spins
or other environmental degrees of freedom in the same manner as for a spin
embedded in a macroscopic solid. For $\Delta/\hbar > 10^{9}$s$^{-1}$ the
parameter $r$ will be small. However, $\alpha$ will be very small compared
to one, and, thus, in accordance with Fig.\ \ref{Fig-Resonance}, no
decoherence of spin oscillations due to coupling with the mechanical
oscillations of the paddle will occur in this case either. The same will be
true even if instead of the paddle one considers, e.g., a Mn$_{12}$ molecule
attached to a carbon nanotube \cite{BogWer08natmat}. The relevant moment of
inertia is now that of the molecules itself, which for a nanometer size
molecule is of order $10^{-42}$kg$\cdot$m$^2$. The corresponding $\omega_r$
is of order $10^{12}$s$^{-1}$ and $\lambda \sim 0.1$. The two regimes are
now $r \gg 1$ for $\Delta/\hbar \ll 10^{12}$s$^{-1}$ and $r \ll 1$, $\alpha
\ll 10^{-2}$ for $\Delta/\hbar \gg 10^{12}$s$^{-1}$. In both limits the
mechanical oscillations should have little effect on coherent spin
oscillations with frequency $\Delta/\hbar$.

The above estimates show that the effects on tunnel splitting and spin
decoherence arising from spin-rotation coupling should not be much of a
concern in nanomechanical setups with large magnetic molecules that have
been discussed in literature. To have a significant effect on the tunnel
splitting one should arrive to $\lambda > 1$. This requires much smaller
moments of inertia, that is, molecules much smaller than Mn$_{12}$.
Consider, e.g., a small magnetic molecule of spin $10$ with the moment of
inertia that is one hundred times smaller than that of the Mn$_{12}$
molecule. We shall also assume that it is coupled to the walls with the
torsional rigidity one hundred times smaller than the coupling through a
carbon nanotube. In this case one still gets $\omega_r \sim 10^{12}$s$^{-1}$
but $\lambda > 1$. Now the regime with $r \sim 1$ is achieved at $%
\Delta/\hbar \sim 10^{12}$s$^{-1}$, which corresponds to $\alpha
\sim 1$. In this case one should expect significant decoherence of
spin oscillations. The bottom line is that decoherence due to a
resonator may occur in subnanometer systems but it should not be
pronounced above the nanometer size. This is easy to understand if
one notices that $I_z\omega_r$ in the expression $\lambda =
S\sqrt{2\hbar/I_z\omega_r}$ is the measure of the
``macroscopicity" of the resonator. Consequently, $I_z\omega \sim
\hbar$ needed to achieve large $\lambda$ generally requires a
system of the atomic size. For larger resonators, interaction
between spin and mechanical degrees of freedom reveals itself only
near the resonance. It results in a very interesting quantum
phenomenon that can be observed in experiment: Splitting of the
mechanical mode of the resonator containing a tunneling spin.
Indeed, in our example with a paddle having $\lambda \sim 10^{-2}$
the splitting of the mechanical mode at the resonance can be quite
significant. For $\Delta < \hbar \omega_r$ the resonance will be
achieved at $W/\hbar \sim \omega_r \sim 10^9$s$^{-1}$, which for
$S=10$ corresponds to the magnetic field of order $10$G. For,
e.g., $\Delta/\hbar \sim 10^8$s$^{-1}$, according to Eq.\
(\ref{splitting}), this will provide the splitting in the MHz
range that would be possible to observe if the quality factor of
the resonator exceeds one thousand.

\section{Acknowledgements}

This work has been supported by the U.S. National Science Foundation through
Grant No. DMR-0703639.

\bibliographystyle{eplbib}
\bibliography{chu-own,gar-tunneling,gar-relaxation,gar-oldworks,gar-books,gar-own,gar-magmech,gar-MM-ordering,gar-spin-boson}

\end{document}